# Controllable 0 - π Josephson junctions containing a ferromagnetic spin valve

E. C. Gingrich[1,2*], Bethany M. Niedzielski[1*], Joseph A. Glick[1], Yixing Wang[1†], D.L. Miller[2], Reza Loloee[1], W. P. Pratt, Jr.[1], and Norman O. Birge[1]

[1] Dept. of Physics and Astronomy, Michigan State University, East Lansing, MI 48824, USA

[2] Northrop Grumman Systems Corporation, Baltimore, MD 21240, USA

[*] These authors contributed equally to this work

[†] present address: Seagate Technology, 47010 Kato Rd, Fremont, CA 94538.

**Superconductivity and ferromagnetism are antagonistic forms of order, and rarely coexist. Many interesting new phenomena occur, however, in hybrid superconducting/ferromagnetic systems. For example, a Josephson junction containing a ferromagnetic material can exhibit an intrinsic phase shift of π in its ground state for certain thicknesses of the material.[1] Such "π-junctions" were first realized experimentally in 2001,[2,3] and have been proposed as circuit elements for both high-speed classical superconducting computing and for quantum computing.[4-10] Here we demonstrate experimentally that the phase state of a Josephson junction containing two ferromagnetic layers can be toggled between 0 and π by changing the relative orientation of the two magnetizations. These controllable 0-π junctions have immediate applications in cryogenic memory where they serve as a necessary component to an ultra-low power superconducting computer.[11] Such a fully superconducting computer is estimated to be orders of magnitude more energy-efficient than current semiconductor-based supercomputers.[12] Phase controllable junctions also open up new possibilities for superconducting circuit elements such as superconducting "programmable logic," where they could function in superconducting analogs to field-programmable gate arrays.**

When a superconducting (S) material and a ferromagnetic (F) material are placed in contact with each other, the properties of both materials are modified near the S/F interface. The intriguing nature of this "superconducting proximity effect" in S/F systems arises due to the exchange field in F, which imposes a phase shift on the two electrons of a Cooper pair as they propagate across F. Cooper pairs in conventional superconductors consist of two electrons with equal and opposite momenta and opposite spin. When such a pair crosses the S/F boundary, one electron goes into the majority, or up-spin, band in F and the other goes into the minority, or down-spin, band causing the two electrons to acquire a net center-of-mass momentum $\pm\hbar Q = \pm(\hbar k_F^\uparrow - \hbar k_F^\downarrow)$, where $\hbar k_F^\uparrow$ and $\hbar k_F^\downarrow$ are the Fermi momenta of the majority and minority bands, respectively.[13] Alternatively, one can say that the electron pair correlation function oscillates in F with wavevector Q perpendicular to the S/F interface. In S/F/S Josephson junctions, those oscillations translate into oscillations between 0-junctions and π-junctions as the F-layer thickness is increased.[1-3]

Imagine now a Josephson junction with the structure S/F$_1$/N/F$_2$/S, where F$_1$ and F$_2$ may be different ferromagnetic materials.[14-16] The pair correlation function describing Cooper pairs from the left-hand S accumulates a phase $\phi_1 = Q_1 * d_{F1}$ while traversing F$_1$, where $d_{F1}$ is the thickness of F$_1$. If the magnetization of F$_2$ is parallel to that of F$_1$, then the pair correlation function will accumulate an additional phase $\phi_2 = Q_2 * d_{F2}$ traversing F$_2$. If, however, the magnetization of F$_2$ is antiparallel to that of F$_1$, then the role of majority and minority bands is reversed, and the pair correlation function will acquire the opposite phase, -$\phi_2$. As shown schematically in Figure 1(a), if we choose $\phi_1$ to be close to π/2 and $\phi_2 \leq$ π/2, then when the layers are parallel $\phi = \phi_P = \phi_1 + \phi_2$, putting the junction into the π state, and when the layers are antiparallel $\phi = \phi_{AP} = \phi_1 - \phi_2$, putting the junction into the 0 state.

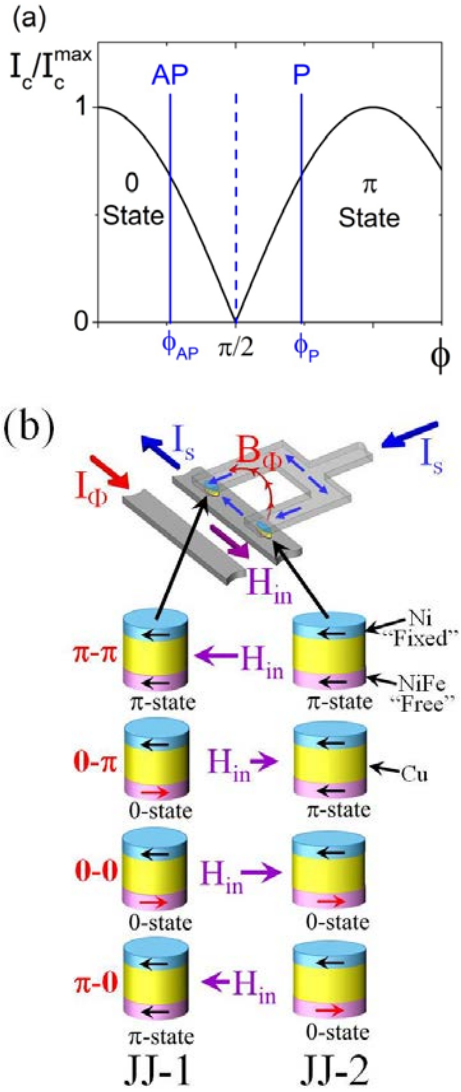

which we accomplish by fabricating a Superconducting QUantum Interference Device, or SQUID, containing two Josephson junctions of the structure described above. The junctions are elliptically shaped with different aspect ratios of 2.2 and 2.8 so that the magnetic layers in the two junctions will have different switching fields. We choose different ferromagnetic materials – one hard for the "fixed layer" and the other soft for the "free layer" – so that only the free layer switches its magnetization direction in small applied magnetic fields. The free layer was chosen as $Ni_{0.80}Fe_{0.20}$ (Permalloy) of thickness 1.5 nm to put the junction close to the 0-π transition [M.A. Khasawneh, BMN, ECG, RL, WPP, & NOB, in preparation]. The fixed layer in the junctions is Ni of thickness 1.2 nm, which should add or subtract a small phase increment.[15,17-19] Figure 1(b) shows a cartoon with the design of our SQUIDs and junctions, as well as the four accessible magnetic states of the junctions. We will use the figure's labeling convention for the four states as "π-π", "0-π", etc. corresponding to the states of the two junctions, JJ-1 and JJ-2 respectively. We will show that these labels accurately describe the phase states of the junctions.

1. **Schematic of experimental design. (a)** Cartoon showing the critical current and phase state of an $S/F_1/N/F_2/S$ Josephson junction as a function of the total phase shift $\phi$ acquired by a Cooper pair traversing the entire structure. The individual phase shifts acquired through $F_1$ and $F_2$ are given in the ballistic limit by $\phi_1 = Q_1 * d_{F1}$ and $\phi_2 = \pm Q_2 * d_{F2}$, respectively. If $\phi_1 = \pi/2$ and $\phi_2 < \pi/2$, then the Josephson junction will be in the π-state when the magnetizations of $F_1$ and $F_2$ are parallel (P), or in the 0-state when the magnetizations are antiparallel (AP). So by controlling the magnetic configuration of the layers in the junction, we can switch a junction between the 0 and π state. **(b)** Schematic diagram of the SQUID, and cartoons showing the magnetization directions of the free and fixed layers for the four magnetic states discussed in this work. The 5 μm wide straight bottom lead, two Josephson junctions and 5 μm wide pitch-fork shaped top lead make up the SQUID device while the 10 μm wide adjacent straight long wire injects magnetic flux Φ into the SQUID loop. The inner dimensions of this loop are 10 μm by 10 μm. The positive directions of various experimental quantities are labeled by arrows: the measurement current, $I_s$, the applied in-plane magnetic field, $H_{in}$, the flux-line current, $I_\Phi$, and the magnetic field produced by the flux line, $B_\Phi$. The more circular junction is labeled as "JJ-1" while the more eccentric elliptical junction is labeled as "JJ-2." The sizes of the four arrows in the four magnetic states depict the magnitude and direction of $H_{in}$ required to reach each state.

Experimental verification of the prediction outlined above requires performing a phase-sensitive measurement,

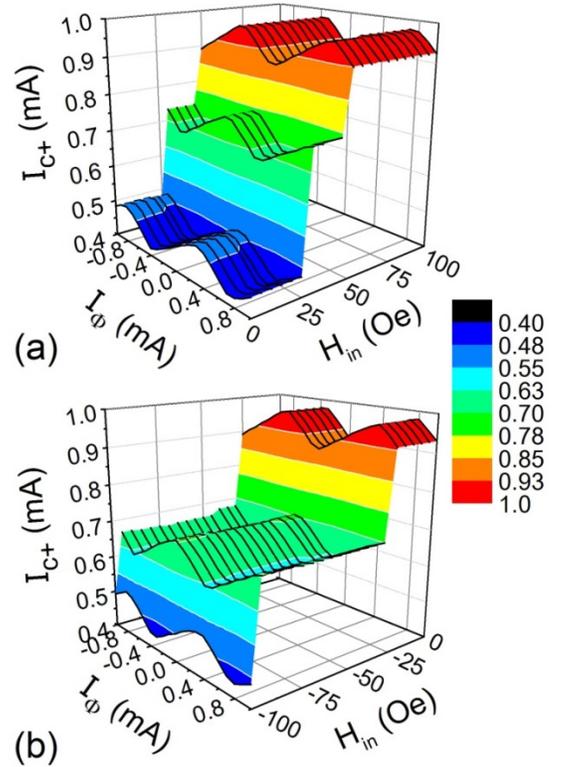

2. **Three-dimensional plots of positive SQUID critical current, $I_{c+}$, vs flux-line current $I_\Phi$ and in-plane set field, $H_{in}$.** After each value of $H_{in}$ is applied, the field is returned to zero and a scan of $I_{c+}$ vs $I_\Phi$ is acquired. Cross-sections at fixed $H_{in}$ show clear SQUID oscillations in $I_{c+}(I_\Phi)$ with a period of about 1.1 mA, corresponding to one flux quantum $\Phi_0 = h/2e$. Sudden jumps in the magnitude and phase of $I_{c+}$ indicate changes in the magnetic state of one of the Josephson junctions in the SQUID. The four total jumps cover the four magnetic states shown in the cartoon of Figure 1(b). **(a)** Data for $H_{in} > 0$. **(b)** Data for $H_{in} < 0$. In both cases the data are taken with $|H_{in}|$ increasing in time.

We initialize the junctions into the π-π state by applying a large in-plane field of $H_{in}$ = -2600 Oe, which sets all four magnetic layers in the negative direction. We then measure, at zero field, a set of I-V curves with different values of the current $I_\Phi$ through the flux line to observe oscillations in the SQUID critical current as a function of applied flux Φ. Critical currents are obtained by fitting I-V curves to the standard form for an overdamped Josephson junction.[20] Note that the critical currents for the two polarities of applied current, $I_{c+}$ and $I_{c-}$, need not be the same. Next we apply a small "set" field $H_{in}$ = 5 Oe, return the field to zero, and repeat the scan of I-V curves vs flux. We continue taking small steps in $H_{in}$, each time setting the field back to zero and repeating a full flux scan. Figure 2(a) shows a 3-dimensional plot of $I_{c+}$ vs $H_{in}$ and $I_\Phi$ as $H_{in}$ is stepped from 0 to 100 Oe. Cross-sections of the plot at fixed values of $H_{in}$ exhibit clear oscillations in $I_c^+(I_\Phi)$. As $H_{in}$ is varied, those cross-sections exhibit two large jumps, the first at $H_{in}$ = 30 Oe and the second at $H_{in}$ = 50 Oe. At each jump, the overall magnitude of the critical current changes, and the peaks in $I_{c+}$ shift along the flux axis. We identify the first jump with the NiFe free layer in the more circular JJ-1 switching its magnetization direction, so that the phase state of JJ-1 switches from π to 0, thus changing the SQUID from the "π-π" to the "0-π" state. The second jump signifies that the NiFe layer in the more elliptical JJ-2 has switched its magnetization direction, and is now also in the 0-state, so the SQUID is now in the "0-0" state. Figure 2(b) shows similar data acquired for $H_{in}$ < 0. Again there are two jumps in the plot, the first occurring at $H_{in}$ = -35 Oe, putting the SQUID in the "π-0" state, and the second at $H_{in}$ = -100 Oe, returning the SQUID to the "π-π" state as at initialization. Taken together, Figures 2(a) and (b) corresponds to a major loop through all four accessible magnetic states of the system. The fact that the magnitudes of the switching fields for $H_{in}$ < 0 are generally larger than for $H_{in}$ > 0 is due to dipolar coupling between the fixed Ni layer and the free NiFe layer in each junction.

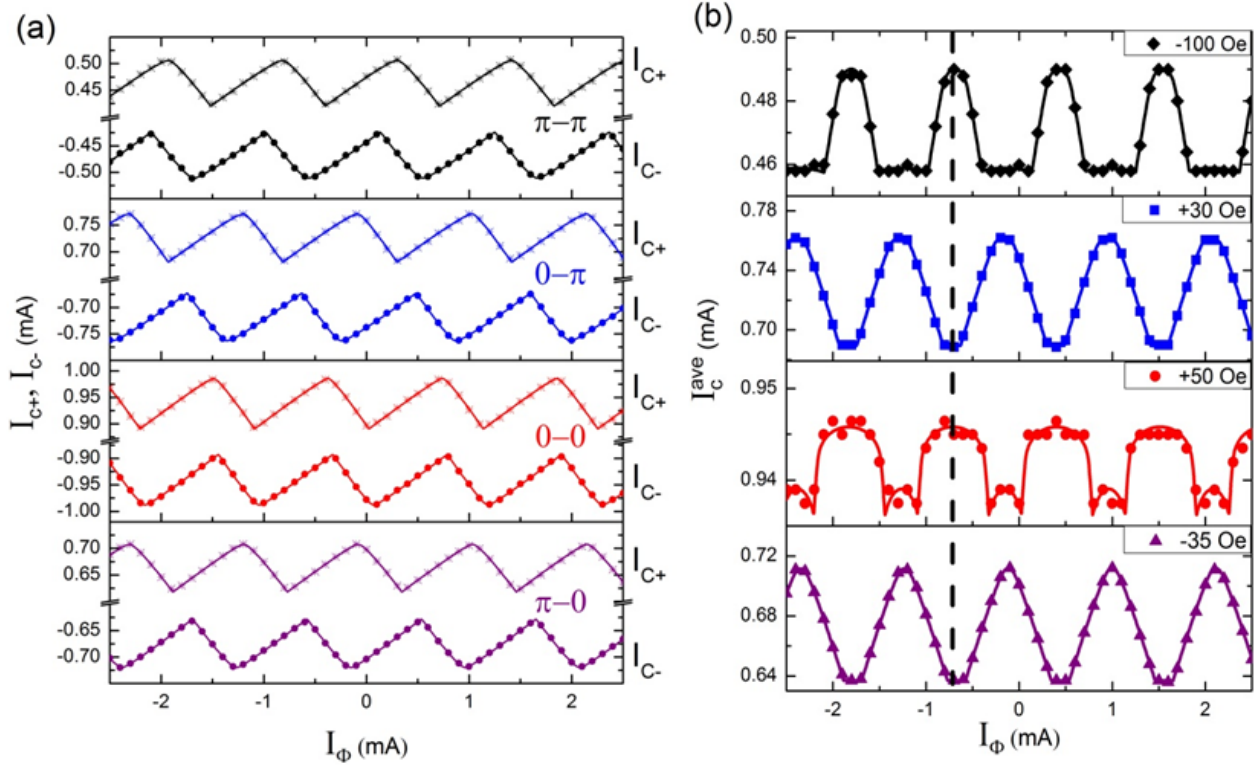

**3. $I_{c+}$, $I_{c-}$, and $I_c^{ave}$ data with fits for all four magnetic states. (a)** Detailed plots of positive and negative SQUID critical currents, $I_{c+}$ and $I_{c-}$, vs flux-line current $I_\Phi$, for the four magnetic states implicated in Figure 2. The states are labeled (π-π), (0-π), etc. according to the phase states of JJ-1 and JJ-2, respectively. $I_{c+}$ and $I_{c-}$ both oscillate as a function of $I_\Phi$, but with a ratchet shape due to the finite and unequal geometrical inductances of the two arms of the SQUID loop. For each magnetic state, the two curves are shifted with respect to each other in opposite directions by amounts that depend on the individual critical currents, $I_{c1}$ and $I_{c2}$, of the two Josephson junctions. These critical currents will change depending on whether the junction is in the 0 or π state. The solid lines are the result of least-squares fits to the data using the asymmetric SQUID model shown in Figure 4, as described in the Supplementary Material. **(b)** Plot of average critical current, $I_c^{ave}$ = $(I_{c+} - I_{c-})/2$ vs $I_\Phi$, for the same four magnetic states represented in (a). The solid lines are derived from the fits in (a). While the shapes of the $I_c^{ave}$ curves depend on the alignment between the $I_{c+}$ and $I_{c-}$ curves, the positions of the maximum and minima in $I_c^{ave}$ are immune to the shifts in $I_{c+}$ and $I_{c-}$. This figure shows schematically the π phase shifts in the (0-π) and (π-0) states relative to the (π-π) and (0-0) states. The analysis presented in the Supplementary Material provides unambiguous proof of the π phase shifts.

Figure 3(a) shows more detailed data of $I_{c+}$ and $I_{c-}$ vs $I_\Phi$ for four selected values of $H_{in}$ taken just after each jump. Several features are immediately apparent in the data. First, $I_{c+}$ and $I_{c-}$ never approach zero; but rather oscillate with an amplitude of about 85 µA in all four magnetic states. Second, the oscillations of $I_{c+}$ and $I_{c-}$ are not sinusoidal, but rather have an asymmetric saw tooth or ratchet shape. Third, the maxima in the $I_{c+}$ and $I_{c-}$ data do not line up with each other, so in general $I_{c-}(\Phi) \neq -I_{c+}(\Phi)$. All three of these features are well understood;[20,21] the first is due to the finite geometrical inductance of the SQUID loop, while the second and third are due to asymmetries in the inductances of the two arms of the loop and in the critical currents of the two junctions.

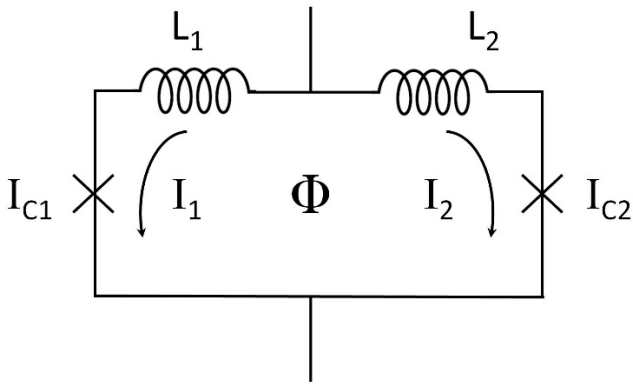

4. **Basic model of asymmetric SQUID**. $I_1$ and $I_2$ are the currents flowing through the two arms, $L_1$ and $L_2$ are the effective inductances of the two arms, and $I_{c1}$ and $I_{c2}$ are the critical currents of the two Josephson junctions. The externally-applied flux through the SQUID is $\Phi$.

A simple model of an asymmetric SQUID is shown in Figure 4 where $L_1$ and $L_2$ are the effective inductances of the two arms of the SQUID loop and $I_1$ and $I_2$ are the currents through each arm.[21,22] Our SQUIDs have an inductance asymmetry, i.e. $L_1 \neq L_2$, because the current paths through the two sides of the SQUID have different lengths (see Figure 1(b)). Our SQUIDs also have an asymmetry in the junction critical currents since the critical current is different when a junction is in the 0 vs the $\pi$ state. Asymmetries in the SQUID loop inductances and in the critical currents of the two junctions cause horizontal shifts of the $I_{c+}(I_\Phi)$ and $I_{c-}(I_\Phi)$ data in opposite directions, which change when the critical current in one of the junctions changes. One can remove those shifts from the data by plotting the average magnitude of the critical current, $I_c^{ave}$ = $(I_{c+} - I_{c-})/2$ vs $I_\Phi$, as shown in Figure 3(b) for the four magnetic states represented in Figure 3(a). The $I_c^{ave}(I_\Phi)$ curves have a variety of shapes depending on how much the $I_{c+}(I_\Phi)$ and $I_{c-}(I_\Phi)$ curves in Figure 3(a) are shifted with respect to each other. Regardless of the shapes, Figure 3(b) shows that the locations of the minima and maxima in $I_c^{ave}(I_\Phi)$ line up with each other, with phase shifts of $\pi$ between successive curves. Figure 3(a) also shows independent fits to the $I_{c+}$ and $I_{c-}$ data, described in the Supplementary Material, which confirm a $\pi$ phase shift between each magnetic state. This demonstrates that we have been able to successfully control the phase of our junctions as proposed above.

The results represented in Figure 3 are reproducible upon repeating the whole major loop. In addition, one can obtain "minor loop" data after initialization by keeping $H_{in}$ between +30 Oe and -35 Oe, so that only the free layer of JJ-1 switches its state. We have obtained similar minor loop data from several different devices; the best major loop data were obtained in the device shown here.

In conclusion, we have demonstrated unequivocally a Josephson junction whose ground state can be switched between the 0-state and $\pi$-state by reversing the magnetization direction of one magnetic layer contained within the junction. Transitions between these states were verified by detecting the additional phase of the $\pi$-state junction within a DC SQUID. These phase controllable junctions have ready application as memory elements in an ultra-low power fully superconducting computer[12] and a design for addressing such a memory has been proposed.[23] The case has been made to use a Josephson junction containing a spin-valve as a memory bit by controlling either the phase or critical current magnitude of the junction.[14-16] Controlling the phase of the junction is advantageous because the magnetic junction can act as a passive phase shifter allowing it to remain in the zero-voltage state during the memory read operation.[11,23] This leads to faster read speeds than when controlling the critical current magnitude.[24] Looking beyond the immediate horizon, one can now start to envision new types of superconducting circuits containing elements with a controllable phase drop, made possible by phase-switching Josephson junctions such as those described here. This should open up new horizons in the nascent field of "superconducting spintronics."[25]

**Acknowledgments:** We thank A. Herr, R. Kleiner, A. Miklich, N. Newman, and N. Rizzo for helpful discussions, J. Willard for performing the FastHenry simulations, and B. Bi for help with fabrication using the Keck Microfabrication Facility. Preliminary work by Y.W. was supported by DOE-BES under grant #DE-FG02-06ER46341. This research is supported by the Office of the Director of National Intelligence (ODNI), Intelligence Advanced Research Projects Activity (IARPA), via U.S. Army Research Office contract W911NF-14-C-0115. The views and conclusions contained herein are those of the authors and should not be interpreted as necessarily representing the official policies or endorsements, either expressed or implied, of the ODNI, IARPA, or the U.S. Government.

**Author Contributions:** E.C.G. and B.M.N. contributed equally to this work. The samples were designed by N.O.B., Y.W. and E.C.G. and fabricated by E.C.G. with significant help from B.M.N. and R.L. The data were taken by B.M.N. using apparatus designed by W.P.P. and software written by Y.W. and E.C.G. The data were analyzed by N.O.B., W.P.P., D.L.M., and J.A.G. The paper was written by N.O.B. and B.M.N with help from W.P.P. and input from all authors. The study was conceived by N.O.B. and W.P.P.

**Author Information:** Correspondence and requests for materials should be addressed to N.O.B. (birge@pa.msu.edu).

# Methods

**Sample Fabrication**

The Josephson junctions and SQUIDs used in this work are fabricated using UHV sputtering deposition and standard microfabrication techniques, including photolithography, e-beam lithography, and ion milling. The bottom wiring layer is a [Nb/Al] multilayer chosen to have less surface roughness than pure Nb, thereby improving the magnetic switching properties of the soft magnetic materials in the junctions.[26] This bottom wiring layer and all of the ferromagnetic layers inside the junction, including a 20-nm top layer of Nb and a final layer of Au to prevent oxidation, are deposited in a single sputtering run without breaking vacuum, to ensure high-quality interfaces. The sputtering chamber is equipped with a liquid-nitrogen-cooled Meissner trap to reduce the partial pressure of water vapor. The base pressure of the sputtering chamber before deposition is $< 2 \times 10^{-8}$ Torr, while the Ar pressure during sputtering is about 2 mTorr. Measurements of the area-resistance product in the normal state yield consistent values of $AR_N \approx 6$ f$\Omega$-m for many junctions of different sizes -- an indicator of the reproducible high quality of the interfaces. The bottom layer is patterned using photolithography and the lift-off process. The junctions are patterned by electron-beam lithography and Ar ion milling, using the negative e-beam resist ma-N2401 as the ion mill mask. The junctions are sufficiently small, with an area of 0.5 $\mu$m$^2$ to ensure that the magnetic layers are single domain.[27] After milling, a SiO$_x$ layer is deposited by thermal evaporation to electrically isolate the junction and the bottom wiring layer from the top wiring layer. Finally, the top Nb wiring layer is deposited by sputtering, again using photolithography and lift-off to define the pitchfork-like pattern seen in Figure 1b). The final SQUID loop has inner dimensions of 10 $\mu$m x 10 $\mu$m with 5 $\mu$m wide strips.

**Measurement**

The measurements reported here were performed at 4.2 K with the samples immersed in a liquid helium dewar equipped with a Cryoperm magnetic shield. The sample dip-stick is equipped with a commercial rf SQUID that is used in a self-balancing potentiometer circuit to measure the voltage across the sample SQUID, and a superconducting solenoid to apply uniform fields in the plane of the sample. The measurement current through the sample SQUID is provided by a battery-powered ultra-low-noise programmable current source. The flux current, $I_\Phi$, is provided by a Yokogawa programmable voltage source and a 1 k$\Omega$ resistor. I-V curves are obtained by sweeping the measurement current I from 0 to a value just above where the sample goes into the voltage state, to determine $I_{c+}$, and then sweeping from 0 in the negative direction to a value just beyond $I_{c-}$.

Initialization of the Ni magnetizations requires applying a large in-plane field $|H_{in}|$ = 2600 Oe. After returning the field to zero, we lift the dip-stick several inches until the sample is just above the liquid helium level and the Nb wiring layers are no longer superconducting, to remove any trapped magnetic flux from the superconducting layers. The sample is then lowered back into the liquid helium and left there for the remainder of the run. The maximum field applied after that is $\pm$100 Oe, which is small enough not to induce any trapped flux in the Nb lines.

## Supplementary Information for
## Controllable 0 - π Josephson junctions containing a ferromagnetic spin valve

E. C. Gingrich[1,2], Bethany M. Niedzielski[1], Joseph Glick[1], Yixing Wang[1], D.L. Miller[2], Reza Loloee[1], W. P. Pratt, Jr.[1], and Norman O. Birge[1]

**Supplementary Discussion**

The standard model for an asymmetric dc SQUID is shown in Figure 4 of the paper. The SQUID is characterized by the four parameters: $L_1$, $L_2$, $I_{c1}$, and $I_{c2}$, which are the effective inductances of the two arms and the critical currents of the two Josephson junctions. $L_1$ and $L_2$ are simply related to the geometric inductances of the two arms if the mutual inductance between them is properly taken into account.[22] Based on the geometry shown in Fig. 1(b), it is expected that $L_2 > L_1$. In our samples $L_1$ and $L_2$ are fixed, whereas $I_{c1}$ and $I_{c2}$ change depending on whether the corresponding junction is in the 0 or $\pi$ state. The externally applied flux is $\Phi$; positive $\Phi$ points out of the page. In addition, the SQUID acquires an extra phase shift of $\pi$, or equivalently an extra flux of $\Phi_0/2$, when one of the two Josephson junctions is in the $\pi$ state. When both junctions are in the $\pi$ state, the two additional phase shifts cancel.

Figure 3(b) of the paper shows the average critical current, $I_c^{ave} = (I_{c+} - I_{c-})/2$, vs current through the flux line, $I_\Phi$. The locations of the maxima and minima in $I_c^{ave}(I_\Phi)$ line up with each other, with phase shifts of $\pi$ between successive curves. Figure 3(b) alone is not sufficient to deduce that the SQUID has acquired an extra phase shift of $\pi$ each time the state of one junction changed, since apparent $\pi$ shifts in the $I_c^{ave}(I_\Phi)$ curves can arise purely from changes in $I_{c1}$ or $I_{c2}$. One can detect the presence of $\pi$ phase shifts by analyzing the shifts in the peak positions of the $I_{c+}$ and $I_{c-}$ curves when the SQUID transitions from state to state, as shown by the following argument. The current reaches its maximum value, $I_c = I_{c1} + I_{c2}$, when the phase drop across each junction is $\pi/2$, so that the currents $I_1$ and $I_2$ through the two arms of the SQUID equal $I_{c1}$ and $I_{c2}$, respectively. Those currents induce a flux through the SQUID loop equal to $\Phi_{self}^+ = L_1 I_{c1} - L_2 I_{c2}$. That must be balanced by the externally-applied flux, so the peak in $I_{c+}$ occurs at a flux of $\Phi_{peak}^+ = -\Phi_{self}^+ = L_2 I_{c2} - L_1 I_{c1}$. The maximum negative critical current occurs at flux $\Phi_{peak}^- = -\Phi_{peak}^+$, hence the peaks in $I_{c+}(\Phi)$ and $I_{c-}(\Phi)$ are shifted with respect to each other by $\Delta\Phi_{peak} \equiv \Phi_{peak}^+ - \Phi_{peak}^- = 2(L_2 I_{c2} - L_1 I_{c1})$. Since the $I_c(\Phi)$ curves are periodic, $\Delta\Phi_{peak}$ can be determined only modulo $\Phi_0$. That means that the center of the pattern – i.e. the point half-way between an $I_{c+}$ peak and its corresponding $I_{c-}$ peak – can only be determined modulo $\Phi_0/2$. Fortunately, it is possible to determine which of the possible values for $\Delta\Phi_{peak}$ is the physical value, by analyzing the changes in $\Delta\Phi_{peak}$ when the critical current of one of the junctions changes. If JJ-1 changes its critical current by an amount $\delta I_{c1}$ while $I_{c2}$ remains unchanged, then the peak separation will change by $\delta(\Delta\Phi_{peak}) = -2L_1 \delta I_{c1}$. Turning the argument around, one can extract the inductance $L_1$ from the transition using $L_1 = -\delta(\Delta\Phi_{peak})/2\delta I_{c1}$. Similarly, if JJ-2 changes its critical current by an amount $\delta I_{c2}$ while $I_{c1}$ remains unchanged, then the peak separation will change by $\delta(\Delta\Phi_{peak}) = +2L_2 \delta I_{c2}$. What one finds is that, if one takes the wrong value of $\delta(\Delta\Phi_{peak})$ for a transition, then the value of $L_1$ or $L_2$ extracted from the transition is unphysical. Our data set provides four transitions; the transitions from the ($\pi$-$\pi$) state to the (0-$\pi$) state and from the (0-0) state to the ($\pi$-0) state allow us to extract $L_1$ since only $I_{c1}$ changes, while the transitions from the (0-$\pi$) state to the (0-0) state and from the ($\pi$-0) state to the ($\pi$-$\pi$) state allow us to extract $L_2$ since only $I_{c2}$ changes. Only one set of phase shifts produces a consistent set of values, $L_1 = 6.0 \pm 0.5$ pH and $L_2 = 11.7 \pm 0.1$ pH, which are listed in the middle row of Table 1.  These values are in agreement with the results of simulations of our SQUID geometry using the FastHenry software, which produced values in the range of 6 – 7 pH for $L_1$ and 13 pH for $L_2$, with variations of about 1 pH for each depending on the type of mesh used in the simulation.  Different choices of $\delta(\Delta\Phi_{peak})$ that don't include the $\pi$ shifts yield values of $L_1$ and $L_2$ that differ in either direction by about 5 pH; those values are not only inconsistent with simulations of our geometry, but more importantly they are inconsistent with the observed depth of the $I_c(\Phi)$ oscillations, which must be approximately equal to $\Phi_0/L$.  The phase shifts deduced from our analysis confirms that the SQUID does indeed acquire a phase shift of $\pi$ each time the system transitions between successive states in the sequence.

Analyzing the four SQUID transitions also provides us with values for $\delta I_{c1}$ and $\delta I_{c2}$, but not values of the critical currents for each junction in each magnetic state.  To estimate the latter, one can assume that the ratios of critical current densities between the 0 and $\pi$ states in both junctions are equal.  That assumption implies that $I_{c1}^0/I_{c1}^\pi = I_{c2}^0/I_{c2}^\pi$, but allows the areas of the two junctions to differ.  That analysis leads to the approximate values $I_{c1}^0 = 560$ µA, $I_{c1}^\pi = 290$ µA, $I_{c2}^0 = 420$ µA, and $I_{c2}^\pi = 220$ µA, with uncertainties of order 10 µA.  The middle row of Table 1 summarizes the results obtained by the preceding analysis.

**Table 1: SQUID and junction parameters**

|  | $L_1$ (pH) | $L_2$ (pH) | $I_{c1}^0$ (µA) | $I_{c1}^\pi$ (µA) | $I_{c2}^0$ (µA) | $I_{c2}^\pi$ (µA) |
|---|---|---|---|---|---|---|
| Preliminary analysis | $6.0 \pm 0.5$ | $11.7 \pm 0.1$ | $560 \pm 10$ | $290 \pm 10$ | $420 \pm 10$ | $220 \pm 10$ |
| Least-squares fits | $5.68 \pm 0.05$ | $11.46 \pm 0.12$ | $565.9 \pm 1.4$ | $292.8 \pm 1.2$ | $419.5 \pm 0.2$ | $210 \pm 7$ |

The inductances of the two arms of the SQUID and the critical currents of each junction in each of its two possible magnetic states are estimated from a simple analysis described in the text (upper row), or by least-squares fitting of the data shown in Figure 3(a) by numerical analysis of the model shown in Figure 4 (lower row).  The much larger uncertainty on the value of $I_{c2}^\pi$ relative to the other three critical currents is a consequence of the fact that $I_{c2}^\pi$ changed slightly when the SQUID transitioned from the ($\pi$-$\pi$) to the (0-$\pi$) state.

To obtain more accurate values of the SQUID parameters, we performed a nonlinear least-squares fit to the data of a numerical analysis of the asymmetric SQUID using the Mathematica software.  The model is described by equations (2.64), (2.72), and (2.73) in [21].  Those equations use the following set of dimensionless variables to describe the SQUID.  The total SQUID inductance is characterized by $\beta_L = LI_c/\Phi_0$, where $L = L_1 + L_2$ is the total loop inductance of the SQUID and $I_c = I_{c1} + I_{c2}$ is the maximum critical current of the SQUID.  If $\beta_L \ll 1$, $L_1 = L_2$, and $I_{c1} = I_{c2}$, then the oscillations of critical current with respect to flux have the standard form, $I_c(\Phi) = I_c|\cos(2\pi\Phi/\Phi_0)|$.  If $\beta_L > 1$, as is the case with our SQUIDs, then the $I_c(\Phi)$ oscillations do not extend to zero, but rather have an amplitude approximately equal to $\Phi_0/L$.  The asymmetry in the inductances of the two arms is characterized by $\alpha_L = (L_2 - L_1)/(L_2 + L_1)$.  As $\alpha_L$ increases, $I_{c-}(\Phi) \neq -I_{c+}(\Phi)$, and the $I_c(\Phi)$ oscillations become more asymmetric, taking on a "ratchet shape" as shown by the data of Figures 2 and 3(a).  The asymmetry in the critical currents is characterized by $\alpha_I = (I_{c2} - I_{c1})/(I_{c2} + I_{c1})$.  As $\alpha_I$ deviates more from zero, the magnitudes of the slopes of both the rising and falling portions of the $I_c(\Phi)$ curves decrease, hence the depth of the $I_c$ modulation decreases. In our samples, $\alpha_L$ should remain the same for all four magnetic states, whereas $\beta_L$ and $\alpha_I$ change because the critical currents change. In addition to determining the shapes of the $I_c(\Phi)$ curves, nonzero values of $\beta_L$, $\alpha_L$, and $\alpha_I$ also shift the positions of the peaks in $I_c(\Phi)$ as discussed above: $\Phi_{peak}^+ = (L_2 I_{c2} - L_1 I_{c1}) = \beta_L (\alpha_L + \alpha_I) \Phi_0/2$.

To reduce the number of free parameters in the fits, we first determined the conversion from $I_\Phi$ to $\Phi$ by fitting the $I_c^{ave}(\Phi)$ curves shown in Figure 3(b) with a simple Fourier Series. For the (0-$\pi$) and ($\pi$-0) states, a single cosine wave fit the data well, while for the (0-0) and ($\pi$-$\pi$) states we used a cosine wave plus its second harmonic. From the Fourier Series fits to the four data sets we determined that $I_\Phi$ = (1115 ± 2)$\mu A * \Phi/\Phi_0$. That conversion factor was then kept fixed in all the ensuing fits. The fitting procedure was as follows. For each magnetic state, the data for $I_c^+(I_\Phi)$ and $I_c^-(I_\Phi)$ were fit simultaneously. The free parameters in each fit were $I_{c+}$, $|I_{c-}|$, $\beta_L$, $\alpha_L$, $\alpha_I$, and $\phi_{shift}$. While we expect $I_{c+}$ and $|I_{c-}|$ to be equal to each other, the data exhibited small differences between $I_{c+}$ and $|I_{c-}|$, of order a few $\mu A$. The last parameter is the shift of the center of the pattern relative to zero flux, $\phi_{shift} \equiv \Phi_{shift}/\Phi_0$. Normally we would expect to observe $\phi_{shift}$ = 0 modulo 1 for the (0-0) and ($\pi$-$\pi$) states, and $\phi_{shift}$ = 0.5 modulo 1 for the ($\pi$-0) and (0-$\pi$) states. The values of the fitting parameters from the fits to all four magnetic states are shown in Table 2. Values for the physical SQUID parameters, $L_1$, $L_2$, $I_{c1}$, and $I_{c2}$ are shown in Table 3. The parameter values shown in the bottom row of Table 1 are averages of the parameters given in Table 3. These tables show that the independent fits to the four data sets provide remarkably consistent results.

**Table 2: Fitting parameters for the four magnetic states of our SQUID.**

| State | $\beta_L$ | $\alpha_L$ | $\alpha_I$ | $I_{c+}$ ($\mu A$) | $|I_{c-}|$ ($\mu A$) | $\phi_{shift}$ | $\phi_{shift} - \phi_{global}$ |
|---|---|---|---|---|---|---|---|
| ($\pi$-$\pi$) | 4.207±0.023 | 0.330±0.006 | -0.147±0.006 | 506.0±0.2 | 510.7±0.3 | -0.1195±0.0007 | 0.0076 |
| (0-$\pi$) | 6.296±0.030 | 0.335±0.005 | -0.472±0.005 | 771.2±0.2 | 762.8±0.2 | +0.3517±0.0006 | 0.4788 |
| (0-0) | 8.198±0.033 | 0.345±0.004 | -0.150±0.004 | 985.5±0.2 | 987.8±0.2 | -0.1360±0.0005 | -0.0089 |
| ($\pi$-0) | 5.957±0.020 | 0.339±0.004 | +0.176±0.004 | 707.7±0.2 | 719.6±0.2 | +0.3955±0.0004 | 0.5226 |

Columns 2 – 6 contain the fitting parameters obtained from the fits to the data for the four SQUID states listed in column 1. The uncertainties are given by the Nonlinear Least Squares fitting routine in Mathematica, and do not account for correlations between the fitting parameters.[S1] The values in the last column are obtained by subtracting $\phi_{global}$ = -0.1271 from the $\phi_{shift}$ values given in column 7. $\phi_{global}$ is the average of the $\phi_{shift}$ values after subtracting 0.5 from the values for the (0,$\pi$) and ($\pi$,0) states.

**Table 3: Physical SQUID parameters obtained from fitting parameters**

| State | $L_1$ (pH) | $L_2$ (pH) | $I_{c1}^0$ ($\mu A$) | $I_{c1}^\pi$ ($\mu A$) | $I_{c2}^0$ ($\mu A$) | $I_{c2}^\pi$ ($\mu A$) |
|---|---|---|---|---|---|---|
| ($\pi$-$\pi$) | 5.73±0.05 | 11.38±0.08 | | 291.6± 1.5 | | 216.8±1.5 |
| (0-$\pi$) | 5.64±0.04 | 11.33±0.07 | 564.5±1.8 | | | 202.5±1.8 |
| (0-0) | 5.63±0.03 | 11.56±0.06 | 567.3± 2.0 | | 419.3±2.0 | |
| ($\pi$-0) | 5.71±0.03 | 11.56±0.05 | | 294.0±1.3 | 419.7±1.3 | |

Values of the inductances of the two arms of the SQUID and of the critical currents of the two Josephson junctions are obtained from the fitting parameters in Table 2. The uncertainties are calculated by the propagation of errors method, but should not be taken too seriously as noted in Table 2. Better estimates of the uncertainties can be had by comparing the parameter values obtained from the fits to the data in the four different magnetic states.

Several comments regarding $\phi_{shift}$ are in order. The major change in $\phi_{shift}$ is the addition of 0.5 in the (0,$\pi$) and ($\pi$,0) states relative to the (0,0) and ($\pi$,$\pi$) states, which is the major result of this work. This is clear from the last column of Table 2, which shows the values of $\phi_{shift}$ for all four magnetic states after subtracting a global shift, $\phi_{global}$ = -0.1271. The origin of the global phase shift is unknown. The experiments take place with the sample mounted vertically inside a Cryoperm shield that is permanently mounted inside the measurement dewar. Rotating the whole dip-stick inside the fixed shield does not cause any change in the $I_{c+}$ or $I_{c-}$ peak positions, so $\phi_{global}$ is not due to flux in the SQUID from a magnetic field source external to the dewar. Some magnetic flux could become trapped in the sample while it is cooled through the superconducting critical temperature, before the dip-stick reached the position of the shield, but additional measurements with a smaller shield mounted directly on the dip-stick did not eliminate $\phi_{global}$. The data in the last column of Table 2 also show that, after subtracting $\phi_{global}$ from $\phi_{shift}$ the values of $\phi_{shift}$ are not exactly equal to 0 or 0.5, but include extra small shifts of about 0.022 in the (0,$\pi$) and ($\pi$,0) states and less than 0.01 in the (0,0) and ($\pi$,$\pi$) states. These small changes in $\phi_{shift}$ are consistent in relative magnitude and sign with what one would expect from the flux coupled into the SQUID by the NiFe layer magnetizations as they change direction. For example, the (0,$\pi$) state has both NiFe magnetizations pointing inward so they induce positive flux into the SQUID loop, thereby causing the peaks in $I_{c+}$ and $I_{c-}$ to shift toward smaller flux. We have calculated the magnitude of these flux changes and find values several times larger than what is observed in the experiment; we believe that most of the flux is shielded by the wide superconducting electrodes placed directly above and below the junctions, which was not taken into account in the calculation.

To complete the analysis, we turn to the much larger shifts in the peak positions that occur each time the magnetic state of the system changes. An approximate analysis of those shifts was discussed earlier. We utilize the dimensionless flux, $\phi = \Phi/\Phi_0$ for this discussion. For each state, the distance between the peaks in $I_{c+}$ and $|I_{c-}|$ is calculated from the least-squares fits: $\Delta\phi_{peak} = \beta_L(\alpha_L + \alpha_I) = 2(L_2I_{c2} - L_1I_{c1})/\Phi_0$. This quantity is not affected by the small shifts discussed in the previous paragraph, since those involve shifts of the $I_{c+}$ and $I_{c-}$ curves in the same direction. The change in $\Delta\phi_{peak}$ from state to state, called $\delta(\Delta\phi_{peak})$, is shown in the second column of Table 4. If only one junction changes its critical current at each change of state, then $\delta(\Delta\phi_{peak}) = 2L_2\delta I_{c2}/\Phi_0$ if JJ-2 changes state, or $\delta(\Delta\phi_{peak}) = -2L_1\delta I_{c1}/\Phi_0$ if JJ-1 changes. Hence $\delta(\Delta\phi_{peak})$ for the ($\pi$–$\pi$) $\rightarrow$ (0-$\pi$) transition should be equal and opposite to $\delta(\Delta\phi_{peak})$ for the (0-0) $\rightarrow$ ($\pi$-0) transition, and similarly $\delta(\Delta\phi_{peak})$ should be equal and opposite for the (0-$\pi$) $\rightarrow$ (0-0) and ($\pi$-0) $\rightarrow$ ($\pi$–$\pi$) transitions. The data in column 2 of Table 4 obey this relation only approximately. This is because of a small change in $I_{c2}$ that occurs during the ($\pi$–$\pi$) $\rightarrow$ (0-$\pi$) transition, which is evident from the last column of Table 3. This is elucidated in the remaining columns of Table 4. Column 3 shows the changes in the total measured $I_c$ of the SQUID for the transitions indicated in column 1. Columns 4 and 5 show the changes in the individual critical currents of the two junctions, using the data provided in Table 3. Let us extract the values of $L_1$ and $L_2$ following the procedure described earlier: for the first and third transitions when JJ-1 switches, we calculate $L_1 = -\delta(\Delta\phi_{peak})\Phi_0/2\delta I_{c1}$ assuming that $\delta I_{c2} = 0$, and similarly we calculate $L_2 = \delta(\Delta\phi_{peak})\Phi_0/2\delta I_{c2}$ for the second and fourth transitions where JJ-2 switches, assuming that $\delta I_{c1} = 0$. That procedure gives the results in the columns 6 and 7 in the table. The results for $L_2$ are consistent with each other, but the results for $L_1$ are not – again because we have neglected the small change in $I_{c2}$ during the ($\pi$–$\pi$) $\rightarrow$ (0-$\pi$) transition. Only when we include both terms and use $\delta(\Delta\phi_{peak}) = 2(L_2\delta I_{c2} - L_1\delta I_{c1})/\Phi_0$, do we find the consistent values for $L_1$ and $L_2$ shown in the last two columns of Table 4.

**Table 4: Analysis of peak shifts and calculation of SQUID inductance.**

|  | Change in $\Delta\phi_{peak}$ | Change in critical currents from state to state: | | | Calculate L using $\delta I_c$ from junction that switches | | Calculate L taking into account $\delta I_c$ of second JJ | |
|---|---|---|---|---|---|---|---|---|
| State change initial → final | $\delta(\Delta\phi_{peak})$ | $\delta(I_c)$ (μA) | $\delta(I_{c1})$ (μA) | $\delta(I_{c2})$ (μA) | $L_1$ (pH) | $L_2$ (pH) | $L_1$ (pH) | $L_2$ (pH) |
| (π–π) → (0–π) | -1.632 | 257 | 272 | -15 | 6.21 |  | 5.58 |  |
| (0–π) → (0–0) | 2.465 | 221 | 3 | 218 |  | 11.71 |  | 11.79 |
| (0–0) → (π–0) | 1.462 | -272 | -274 | 2 | 5.52 |  | 5.43 |  |
| (π–0) → (π–π) | -2.295 | -206 | -1 | -205 |  | 11.59 |  | 11.62 |

See description in text for the analysis of the shifts in the relative peak positions of the $I_{c+}$ and $I_{c-}$ data.

Finally, one can ask how we can be sure that our fitting procedure "chose" the right values for $\Delta\phi_{peak}$, given that $\Delta\phi_{peak}$ is only determined modulo 1 by the positions of the peaks in $I_{c+}(\phi)$ and $I_{c-}(\phi)$. For example, one could add any integer multiple of $1/\beta_L$ to either $\alpha_L$ or to $\alpha_I$, without changing the value of $\Delta\phi_{peak} = \beta_L(\alpha_L + \alpha_I)$ modulo 1. It turns out that the fits are extremely sensitive to $\alpha_L$, since that parameter is determined by the relative slopes of the rising and falling portions of the $I_{c+}(\phi)$ and $I_{c-}(\phi)$ data. Also, since $\alpha_L$ is a purely geometric parameter, it should be the same for all four magnetic states, so it would be very suspicious if the fitting procedure returned different values of $\alpha_L$ for different states of the system. Very small changes in $\alpha_I$, however, might not cause the fit to deviate significantly from the data points. To rule out this possibility, we have repeated the analysis shown in Table 4 assuming that the (π-0) and (0-π) states do not carry an extra phase shift of π. This is equivalent to adding or subtracting 1 from the values of $\delta(\Delta\phi_{peak})$ in Table 4. Following through this procedure, one finds inductance values that are either much too small ($L_1 \approx 1$ pH and $L_2 \approx 7$ pH) or much too large ($L_1 \approx 10$ pH and $L_2 \approx 16$ pH). As discussed earlier, those values are incompatible with the estimates of the inductances obtained from FastHenry, and, more importantly, with the observed depth of the modulations in $I_{c+}(\phi)$ and $I_{c-}(\phi)$ data.

**References**

S1. Press WH, Flannery BP, Teukolsky SA & Vetterling WT, Ch. 14 of *Numerical Recipes*, Cambridge Univ. Press, Cambridge (1988). Due to the large number of fitting parameters, a complete analysis of correlations in parameter uncertainties is beyond the scope of this work. Carrying out such an analysis would not affect any of the conclusions of this work.